\documentclass[twocolumn,epsfig]{aa}
\usepackage{graphicx}
\begin{document}

\title{Amplification of MHD waves in swirling astrophysical flows}

\author{Andria D. Rogava\inst{1,2}\thanks{On leave from Abastumani
Astrophysical Observatory, Kazbegi ave.~2a,
Tbilisi--380060, Georgia}
   \and Gianluigi Bodo\inst{3}
   \and Silvano Massaglia\inst{1}
   \and Zaza Osmanov \inst{4}}

\institute{Dipartimento di Fisica Generale,
          Universit\'a degli Studi di Torino, Via Pietro Giuria 1, Torino
          I-10125, Italy
\and   Abdus Salam International Centre for Theoretical Physics,
          Trieste I-34014, Italy
\and   Osservatorio Astronomico di Torino, Strada
          dell'Osservatorio 20, I-10025, Pino Torinese, Italy
\and   Centre for Plasma Astrophysics, Abastumani Astrophysical
       Observatory, Tbilisi--380060,  Georgia }
\offprints{Andria Rogava}
\date{Received / Accepted }
\authorrunning{Rogava et al.}
\titlerunning{Amplification of MHD waves in swirling astrophysical flows}

\abstract{Recently it was found that helical magnetized flows
efficiently amplify Alfv\'en waves (\cite{rmbm03}). This robust
and manifold nonmodal effect was found to involve regimes of
transient algebraic growth (for purely ejectional flows), and
exponential instabilities of both usual and parametric nature.
However the study was made in the incompressible limit and
an important question remained open - whether this amplification
is inherent to swirling MHD flows {\it per se} and what is the
degree of its dependence on the incompressibility condition. In
this paper, in order to clear up this important question, we
consider full compressible spectrum of MHD modes: Alfv\'en waves
(AW), slow magnetosonic waves (SMW) and fast magnetosonic waves
(FMW). We find that helical flows inseparably blend these waves
with each other and make them unstable, creating the efficient
energy transfer from the mean flow to the waves. The possible role
of these instabilities for the onset of the MHD turbulence,
self-heating of the flow and the overall dynamics of astrophysical
flows are discussed.}

\maketitle

\section{Introduction}

Lately it became quite commonly presumed and hoped that swirling
three-dimensional motion occurs in different kinds of space plasma
flows. The presumption is grounded on the mounting evidence from
different branches of observational astronomy, while the hope is
related to the long-standing aspiration of theoreticians to find
in astrophysical flows natural laboratories for testing of their
expectations and ideas.

One of the most remarkable observational indications came from
{\it Solar and Heliospheric Observatory---Coronal Diagnostic
Spectrometer} (SOHO-CDS) data that led to the identification
(\cite{pm98}) of macrospicules having both rotational and jet-like
features. Pike and Mason presented evidence for the existence of
blue- and red-shifted emission on either side of an axis of a
macrospicule stretching above the limb from a footpoint region on
the disk. They interpreted these observations as indicating the
presence of a {\it rotation} within these tall squalls of hot and
magnetized, swirling plasma flows --- christened as {\it solar
tornados} --- observed both on the limb and the disk of the Sun.
This discovery allows to argue (for earlier arguments in the same
vein see \cite{po86,su86}) that the wide range of dynamic events
in the solar atmosphere, including micro flares, jets, plumes,
surges, spicules and macrospicules, may exhibit complicated,
kinematically nontrivial, three-dimensional sheared plasma
motions. Generally speaking it looks quite credible that rotation
plays a major role in the dynamics of chromospheric and transition
region features. It seems plausible to admit that the presence of
these complicated patterns of plasma motion deeply influence the
dynamics of MHD waves in the solar atmosphere, contribute to the
coronal heating, and to the acceleration of the solar wind.

Another message of observational evidence for swirling flows
recently came from the totally different class of astronomical
objects: Herbig-Haro (HH) jets. Two different observations of
young stellar objects in HH 212 (\cite{d00}) and DG Tau
(\cite{b02}), contain serious indications for the presence of
rotation in these jets. These results are in accordance with
predictions of the popular magnetocentrifugal jet launching model
(\cite{bp82}). The possible impact of these complicated, helical
motions on physical processes occurring within these flows is yet
to be understood.


The growing evidence in favour of astrophysical helical flows
poses a twofold challenge. At the one hand it naturally solicits
for the further theoretical study of {\it general} aspects ---
generation, equilibrium, stability and internal dynamics --- of
helical flows in the framework of plasma astrophysics. At the
other hand, after acquiring better understanding of basic physics,
it suggests to build {\it concrete} prototype models, closely
adjusted to specific examples of observed swirling flows, which
might appear to be useful for shedding some light on the puzzling
observational appearance of related astronomical objects.

The basic theory of shear flows tells us that dynamics of waves
and vortices, sustained by these flows, are substantially affected
by the differential character of motion. On the basic mathematical
level these phenomena, somewhat misleadingly christened as {\it
``nonmodal processes"}, are related to the non-self-adjointness of
linear dynamics of perturbations in SF (\cite{ttrd93}). The
variety of these processes, that could be labeled as {\it
``shear-induced nonmodal processes"} (SINP), is quite
well-understood for simple, plane parallel flows and thoroughly
described in the recent literature (for the latest review see,
e.g., \cite{bprr01} and references therein). It is known that SINP
lead to the generation of new modes of plasma collective
behaviour, to new forms of flow-wave, wave-wave and vortex-wave
interactions provoked and fueled by `parent' shear flows. These
phenomena, originally disclosed in hydrodynamics, take place in
various kinds of plasmas and might have a number of astrophysical
applications, including pulsar magnetospheric plasmas, solar
atmospheric phenomena and galactic gaseous disk dynamics. The
interest towards possible applications was strengthened by the
recent numerical evidence that real-space appearance of SINP is
easily recognizable and robust even in the presence of tangible
dissipation (\cite{bprr01}).

Real astrophysical shear flows are almost always involved in
motions with complex kinematics/geometry and {\it kinematic
complexity} is known to bring an additional variety to SINP. When
geometry and kinematics of flows are complex a whole bundle of new
effects arises, like ``echoing" (repetitive) transient pulsations
and different kinds of shear-induced (including parametric)
instabilities (\cite{mr99}). The study of kinematically complex SF
was initiated in hydrodynamics (\cite{lfl84,cc86,cd90}) and this
approach is still largely unknown for the plasma and astrophysics
community. Therefore, it is an intriguing challenge and a task of
a big practical importance to study these processes in MHD flows
of nontrivial geometry and kinematics.


The first step in this direction was recently made (\cite{rmbm03},
hereafter referred as Paper I). In this study the simplifying
assumption of incompressibility was adopted, which cuts off modes
of acoustic origin (slow magnetosonic waves, SMW, and fast
magnetosonic waves FMW) and allows to concentrate on the
investigation of the dynamics of Alfv\'en waves (AW). The subject
of the interaction between flows and AW is interesting in a number
of astrophysical applications
(\cite{bh91,tpc92,tp99,rhwt01,vt02}). In Paper I it was found that
helical shear flows are efficient amplifiers of AW. In purely
ejectional flows (i.e., when no rotation is present) AW are
amplified transiently via algebraic, shear-induced instability. In
a swirling flow AW are exponentially unstable: depending on the
mode of differential motion both usual and parametric
instabilities appear.


These results were discussed in the context of their possible
(observable) manifestations. It was argued that they might account
for the generation of the large-amplitude Alfv\'en waves -- e.g.,
within `tornado-like' patterns existing  in the solar atmosphere.
It was suggested that they could lead to the efficient {\it
self-heating} of flows: the kinetic energy of the flow, being
extracted by amplified Alfv\'en waves, returns back to the flow in
the form of the thermal energy because AW are eventually damped
via magnetic diffusion. Finally, it was argued that these
instabilities might serve as an initial (linear and nonmodal)
phase in the ultimate {\it subcritical} transition to MHD
Alfv\'enic turbulence in various kinds of astrophysical shear
flows.

It is known that the incompressibility condition, used in the
Paper I, is quite restrictive. Often the usage of this condition
introduces its own imprints on the dynamics of perturbations and
it is not trivial to distinguish the genuine SINP from the
phenomenological effect imposed by the incompressibility
approximation. Therefore it is quite important to consider MHD
waves in helical flows {\it without} the usage of the
incompressibility condition and to study the SINP for the full
spectra of MHD waves containing together with the AW also the SMW
and the FMW.

This task is undertaken in this paper. We find that the range of
processes, sustained by helical flows is extremely rich. It
encompasses different kinds of wave transformations, wave
beatings. Most importantly, usual and parametric shear
instabilities, which has been found in the Paper I, are found to
appear for the full spectra of MHD waves containing AW, SMW and
FMW. We see that flows of such a high degree of complexity
efficiently intertwine all three MHD wave modes and efficiently
exchange energy with them. The relevance of these results to the
physics of helical flows and to the understanding of possible
observational appearance of related astronomical objects are
pointed out and critically discussed.

\section{Theory}

Our aim is to study linear collective MHD modes in helical flows.
For this purpose we need to write equations of the ideal
linearized MHD for the evolution of perturbations within the flow.
In Paper I we studied only incompressible perturbations, while now
we consider fully compressible case, so our starting equations are
[$D_t{\equiv}{\partial}_t+({\bf V}{\cdot}{\nabla})$]:
$$
D_t{\rho}+{\rho}({\nabla}{\cdot}{\bf V})=0,\eqno(1)
$$
$$
D_t{\bf V}=-{1 \over{{\rho}}}{\nabla}P-{{{\bf B}} \over {4{\pi}{\rho}}}{\times}
({\nabla}{\times}{\bf B}), \eqno(2)
$$
$$
D_t{\bf B}=({\bf B}{\cdot}{\nabla}){\bf V}-{\bf B}({\nabla} {\cdot}{\bf V}),
\eqno(3)
$$
$$
{\nabla}{\cdot}{\bf B}=0. \eqno(4)
$$


Our equilibrium model, used in {\it Paper I}, assumes a
homogeneous MHD plasma ($\rho_0=const$), embedded in a
homogeneous, vertical magnetic field (${\bf B}_0{\equiv}[0,~0,~
B_0=const]$). We consider instantaneous values of all physical
variables as sums of their mean (equilibrium) and perturbational
components: ${\bf B}{\equiv} {\bf B}_0+{\bf B}^{\prime}$,
${\rho}{\equiv}{\rho}_0+{\rho}^{\prime}$, etc. Applying this
decomposition we convert (1--4) into the following set for
perturbation variables [${\cal D}_t{\equiv}{\partial}_t+({\bf
U}_0{\cdot}{\nabla})$]:
$$ {\cal D}_td+{\nabla}{\cdot}{\bf u}=0, \eqno(5)
$$
$$
{\cal D}_t{\bf u}+({\bf u}{\cdot}{\nabla}){\bf
U}_0=-C_s^2{\nabla}d+ C_A^2[{\partial}_z{\bf b}-{\nabla}b_z], \eqno(6)
$$
$$
{\cal D}_t{\bf b}=({\bf b}{\cdot}{\nabla}){\bf
U}_0+{\partial}_z{\bf u} +{\bf e}_z({\nabla}{\cdot}{\bf u}), \eqno(7)
$$
$$
{\nabla}{\cdot}{\bf b}=0, \eqno(8)
$$
with
$d{\equiv}{\rho}^{\prime}/{\rho}_0$ and ${\bf b}{\equiv}{\bf
B}^{\prime} /B_0$. Note also that for compressible perturbations
$p^\prime=C_s^2 \rho^\prime$ with $C_s$ the homogeneous speed of
sound.

If we introduce a new vector quantity ${\bf h}{\equiv}{\bf b}-{\bf
e}_zd$ then the above system can effectively be reduced to the
following set of second-order equations:
$$
{\cal D}_t^2{\bf
h}-[({\bf h}{\cdot}{\nabla}){\bf U}_0{\cdot}{\nabla}]{\bf U}_0-
(C_s^2+C_A^2){\Delta}{\bf h}=
$$
$$
C_A^2[{\partial}_z^2{\bf
h}-{\nabla}({\partial}_zh_z)-{\bf e}_z{\partial}_z
({\nabla}{\cdot}{\bf h})], \eqno(9)
$$
which describes evolution
of all three MHD modes --- SMW, AW and FMW --- influenced by the
presence of the equilibrium flow with an arbitrary ${\bf U}_0$.


In paper I we worked with the velocity field specified by
$$
{\bf U}(r){\equiv}[0,~r{\Omega}(r),~U(r)], \eqno(10)
$$
with ${\Omega}(r)={\cal A}/r^n$, where $r=(x^2+y^2)^{1/2}$ is a
distance from the rotation axis, while ${\cal A}$ and $n$ are some
constants. In particular, $n=0$ and ${\cal A}={\Omega}_0$ for
rigidly rotating plasmas, while $n=3/2$ and ${\cal A}=(GM)^{1/2}$,
for the Keplerian rotation.

The linear shear matrix for this kind of ${\bf U}(r)$ has the following form
({\cite{rmbm03}):
$$
{\cal S}={\left(\matrix{{\sigma}&A_1&0\cr
A_2&-{\sigma}&0 \cr C_1&C_2&0\cr}\right)}. \eqno(11)
$$

The key equation of the ``nonmodal" method (\cite{cc86,mr99}),
$$
\partial_t{\bf k}+{\cal S}^T\cdot{\bf k}=0, \eqno(12a)
$$
giving a full evolutionary picture of ${\bf k}$'s, transcribes
in this case to the following set of equations\footnote{Hereafter
$F^{(n)}$ will denote $n$-th order time derivative of a function
$F$.}:
$$
k_x^{(1)}+{\sigma}k_x+A_2k_y+C_1k_z=0, \eqno(12b)
$$
$$
k_y^{(1)}+A_1k_x-{\sigma}k_y+C_2k_z=0, \eqno(12c)
$$
while $k_z=const$.

These equations
imply that , while $k_x(t)$ and $k_y(t)$ may have algebraic,
exponential or periodic time dependence. For swirling flows
the differential rotation parameter $n$ plays decisive role in determining
the evolution scenario for the wave number vector $|{\bf k}(t)|$: when $n<1$
(including the rigid rotation case) the time
evolution of the $|{\bf k}(t)|$ is periodic, while when $n>1$ (including the
Keplerian rotation regime) $|{\bf k}(t)|$ evolves
exponentially.

Note that the following nonlinear
combination of $k_x(t)$ and $k_y(t)$:
$$
{\Delta}{\equiv}k_xk_y^{(1)}-k_yk_x^{(1)}+k_z(C_1k_y-C_2k_x)=
$$
$$
A_2K_y^2-A_1k_x^2+2{\sigma}k_xk_y+2k_z(C_1k_y-C_2k_x)=const.
\eqno(13)
$$
is a conserved quantity. In the case of pure,
two-dimensional rotation ($k_z=C_1=C_2=0$) it reduces to the
constant ${\Delta}_r{\equiv}k_xk_y^{(1)} -k_yk_x^{(1)}$
(\cite{mr99}), while for the case of a pure outflow ($A_1=A_2=
{\sigma}=0$) it reduces to the conservation of the quantity
${\Delta}_e{\equiv} C_1k_y-C_2k_x$ (\cite{rpm00}). Therefore, we
see that the {\it wave-number invariant} $\Delta$ in the case of
the helical flow is the sum of the ${\Delta}_r$ and the ${\Delta}_e$
functions.

Using the same method as in {\it Paper I} (see for details also \cite{mr99})
we can effectively convert the system (5--8) to the set of first oder ordinary
differential equations [${\bf H}{\equiv}i{\bf h}$, $D{\equiv}id$]:
$$
D^{(1)}={\bf k}{\cdot}{\bf u}, \eqno(14)
$$
$$
{\bf u}^{(1)}+{\cal S}{\cdot}{\bf u}=-(C_s^2+C_A^2){\bf k}D
+
$$
$$
C_A^2[k_z{\bf H}-{\bf k}H_z+ {\bf e}_zk_zD], \eqno(15)
$$
$$
{\bf H}^{(1)}={\cal S}{\cdot}{\bf H}-k_z{\bf u}, \eqno(16)
$$
$$
{\bf k}{\cdot}{\bf H}=-k_zD, \eqno(17)
$$

Note that (14-17) contain two first order
ordinary differential equations with time-dependent coefficients.
Time-dependence of these coefficients is completely determined by the
temporal evolution of the wave number vector ${\bf k}(t)$ and,
therefore, governed by Eq. (12). We can reduce this set to the set of second
order equations for the components of the vector ${\bf H}$, which
is the nonmodal form of the Eq. (9). It can be written in
the following (vector) form:
$$
{\bf H}^{(2)}+C_s^2{\bf k}({\bf k}{\cdot}{\bf H})-{\cal S}^2{\cdot}{\bf H}+
C_A^2[({\bf k}-k_z{\bf e}_z)({\bf k}{\cdot}{\bf H})+
$$
$$
(k_z^2{\bf H}-{\bf k}k_zH_z)]
=0. \eqno(18)
$$

Note that in the helical flow, specified by the shear matrix (11), the square of the matrix
is equal to (\cite{rmbm03}):
$$
||{\cal S}^2||={\left(\matrix{
                               {\Gamma}^2     &           0     & 0 \cr
                               0              & -{\Gamma}^2     & 0 \cr
                               {\varepsilon}_1& {\varepsilon}_2 &0  \cr}
\right)}, \eqno(19)
$$
where we use notation: ${\Gamma}{\equiv}({\sigma}^2+A_1A_2)^{1/2}$,
${\varepsilon}_1{\equiv}A_2C_2+{\sigma}C_1$ and
${\varepsilon}_2{\equiv}A_1C_1-{\sigma}C_2$. Note that $||{\cal S}^3||=
{\Gamma}^2||{\cal S}||$.

Splitting both ${\bf H}$ and ${\bf k}$ vectors into their longitudinal and
transverse components --- ${\bf H}{\equiv}({\bf H}_{\perp},~H_z)$,
${\bf k}{\equiv}({\bf k}_{\perp},~k_z)$ ---
we can write more explicit form of (18) revealing the nature of coupling between
the MHD modes, imposed by the presence of the velocity shear:
$$
H_z^{(2)}+C_s^2k_z^2H_z+C_s^2k_z({\bf k}_{\perp}{\cdot}{\bf H}_{\perp})-
({\varepsilon}_1H_x+{\varepsilon}_2H_y)=0, \eqno(20)
$$
$$
{\bf H}_{\perp}^{(2)}+[C_A^2k_z^2-{\Gamma}^2]{\bf H}_{\perp}+(C_s^2+C_A^2)
{\bf k}_{\perp}({\bf k}_{\perp}{\cdot}{\bf H}_{\perp})+
$$
$$
C_s^2{\bf k}_{\perp}k_zH_z=0. \eqno(21)
$$

It is easy to see that in the absence of the shear flow Eqs. (20-21)
give standard
expressions for the dispersion properties of all these modes. The much simpler
version of this system was analyzed before for
plane-parallel SF of standard MHD plasmas (\cite{crt96}). In
the astrophysical context the similar kind of coupled ODE's were
studied for MHD waves in the solar wind plasmas (\cite{prm98}), in
galactic gaseous discs (\cite{rph99}) and in cylindrical rotationless
flux tubes (\cite{rpm00}). The factual `route'
of the wave
number vector ${\bf k}(t)$  temporal ``drift" (caused, in its turn, by the
differential character of the plasma motion) plays the
crucial role in the time evolution of physical perturbations. In parallel
flows ${\bf k}_{\perp}(t)$ exhibits linear time dependence. However, even
in this relatively simple case we have the whole set of SINP
(\cite{rpm00}): waves exchange energy with the flow and the velocity shear
makes waves coupled with one another, making possible their reciprocal
transformations.

Helical flows are expected to exhibit
evolutionary regimes characteristic to both purely ejectional and
purely rotational flow patterns. However, we expect that shear-induced wave
transformations (SIT) are more distinctly exhibited by parallel flows
without rotation in the transverse
plane, because these flows are stable and SIT is the only major kind of
SINP occurring in the flow. In helical
flows, presumably featuring different sorts of
shear-induced instabilities, SIT could be less- or not-pronounced.
It should be noted that the parameters ${\varepsilon}_1$ and
${\varepsilon}_2$ exist only when both rotation and outflow are
present and they are nonzero only when the forces that determine
the kinematic portrait of the flow are non-conservative
(\cite{cc86}).  Thus their role in the dynamics of MHD waves sustained by
swirling flows can be quite significant.

\section{Discussion}

The evolution of linear MHD modes in kinematically complex helical flows
might be quite complicated. Before considering any special and/or particular
cases we have to recall the following three levels of complexity arising
within this problem.

First, the {\it medium} itself is complex enough, because it sustains three
different linear modes of oscillations: SMW, AW, and FMW. Although
in the absence of shear these modes are decoupled, still an
arbitrary perturbation excited within this medium is normally a
superposition of these three normal modes (\cite{s94}).

Second, in the presence of a simple, plane-parallel flow with a linear
velocity profile:
(a) one mode - FMW - becomes able to
draw energy out of the background flow; (b) depending on the
plasma-$\beta$  different waves become coupled and are able
to transform into each other; (c) the system
starts exhibiting beat wave phenomena. The resulting picture of
the MHD wave dynamics becomes considerably complex (\cite{rpm00}).

Third, when incompressibility condition is used and the presence of the rotation
and stretching of flow lines in the
transverse cross section of the flow is ``allowed" one finds that new kinds of
shear instabilities emerge  (\cite{rmbm03}). This is expected  to happen also
for compressible (acoustic) wave modes, because even in the simplest
two-dimensional, nonmagnetized flow
with kinematic complexity (\cite{mr99}) usual and parametric instabilities
do appear together
with asymptotically persistent, ``echoing" solutions.

Therefore it seems reasonable to suppose that helical MHD flows, possessing
all these levels (or degrees) of complexity plus the complexity of
the specifically helical nature, related with the existence of the
${\varepsilon}_1$ and ${\varepsilon}_2$ ``helical" parameters, must
exhibit highly complicated collective processes, dominated by
different regimes of shear-induced variability and instability.

For numerical purposes it is more suitable to deal with the
dimensionless version of the (14-17) set:
$$
{\varrho}^{(1)}={\cal K}_x(\tau)v_x+{\cal
K}_y(\tau)v_y+ v_z, \eqno(22)
$$
$$
v_x^{(1)}+{\Sigma}v_x+a_1v_y=-{\cal
K}_x(\tau){\epsilon}^2 {\varrho}+b_x- {\cal K}_x(\tau)b_z,
\eqno(23a)
$$
$$
v_y^{(1)}+a_2v_x-{\Sigma}v_y=-{\cal
K}_y(\tau){\epsilon}^2{\varrho}+ b_y- {\cal K}_y(\tau)b_z,
\eqno(23b)
$$
$$
v_z^{(1)}+R_1v_x+R_2v_y=-{\epsilon}^2{\varrho},
\eqno(23c)
$$
$$
b_x^{(1)}={\Sigma}b_x+a_1b_y-v_x,
\eqno(24a)
$$
$$
b_y^{(1)}=a_2b_x-{\Sigma}b_y-v_y,
\eqno(24b)
$$
$$
{\cal K}_x(\tau)b_x+{\cal K}_y(\tau)b_y+b_z=0,
\eqno(25)
$$
derived from (14-17) with the usage of dimensionless
notation: ${\tau}{\equiv}C_Ak_zt$, ${\cal
K}_x(\tau){\equiv}k_x(t)/k_z$, ${\cal
K}_y(\tau){\equiv}k_y(t)/k_z$, $v_i{\equiv}u_i/C_A$, $\Sigma
\equiv \sigma/C_Ak_z$, $a_{1,2}{\equiv}A_{1,2}/C_Ak_z$,
$R_{1,2}{\equiv}(C_{1,2}/k_zC_A)$, ${\epsilon} {\equiv}C_s/C_A$.

It is instructive to calculate how the total energy of
perturbations being, in this case, the sum of kinetic,
compressional and magnetic energies:
$$
E_{\rm tot}{\equiv}E_{\rm kin}+E_c+E_m, \eqno(26)
$$
$$
E_{\rm kin}{\equiv}(v_x^2+v_y^2+v_z^2)/2, \eqno(27a)
$$
$$
E_{\rm c}{\equiv}{\epsilon}^2{\varrho}^2/2, \eqno(27b)
$$
$$
E_{\rm m}{\equiv}(b_x^2+b_y^2+b_z^2)/2, \eqno(27c)
$$
changes in time. We can easily see that the evolutionary equation for
the total energy is:
$$
E^{(1)}=(a_1+a_2)(b_xb_y-v_xv_y)+{\Sigma}[(v_y^2-v_x^2)+(b_x^2-b_y^2)]+
$$
$$
R_1(b_xb_z-v_xv_z)+R_2(b_yb_z-v_yv_z), \eqno(28)
$$

The scope of this paper is not full investigation of all possible regimes
of evolution. Instead, our purpose is to see whether shear instabilities,
disclosed in the incompressibility limit for AW,
appear also in the compressible case, for the blend of SMW, AW, and FMW.
It is instructive and convenient to unfold these phenomena by
following the scheme used in our previous studies of cylindrical
flows (\cite{rpm00} and the Paper I) and to see how two,
major regimes of ${\bf k}(t)$-dynamics (periodic and exponential)
affect the course of collective phenomena in flows with different
values of the plasma-$\beta$.

\subsection{High-$\beta$ plasmas}

In parallel flows, when ${\epsilon}^2{\gg}1$, FMW are decoupled
from the AW and SMW, while the latter two are coupled (\cite{rpm00})
and may transform into each other. Swirling
flows show more complicated behaviour. When ${\gamma}^2{\equiv}
({\Gamma}/C_Ak_z)^2<0$, i.e.,
when temporal evolution of wavenumber vectors $\bf k(t)$ is
periodic, we encounter with two new effects:

\begin{figure}
  \resizebox{\hsize}{!}{\includegraphics[angle=90]{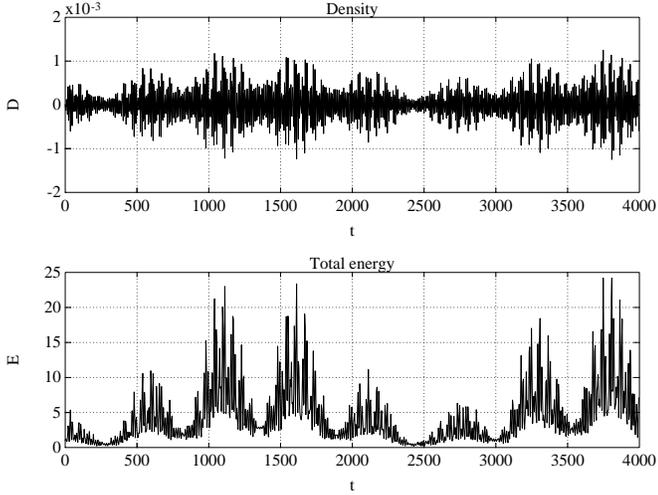}}
  \caption{The temporal evolution of the density $D(\tau)$ and the normalized
  total energy of perturbations $E_{\rm tot}(\tau)/E_{\rm tot}(0)$,
which exhibits quasiperiodic and pulsational behavior. The set of
parameters is: $\epsilon=10$, ${\cal K}_x(0)=10$, ${\cal
K}_y(0)=8$, $R_1=0.8$, $R_2=2$, $\Sigma=0$, $a_1=-0.1$, $a_2=0.1$.
}
\end{figure}

\begin{figure}
  \resizebox{\hsize}{!}{\includegraphics[angle=90]{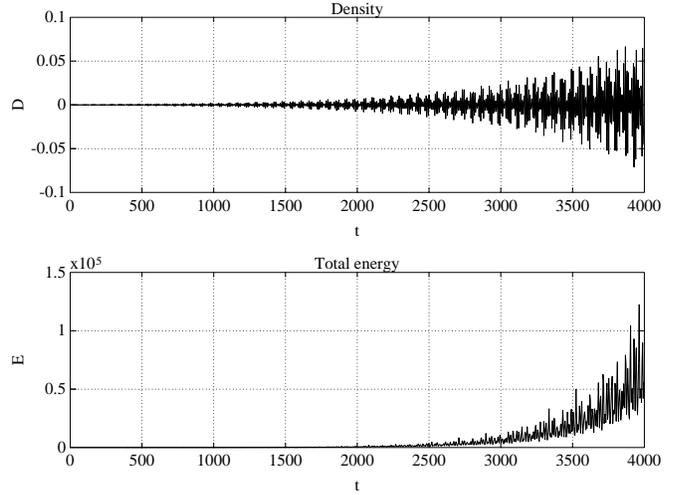}}
  \caption{The evolution of parametrically unstable blend of AW and SMW.
  The set of parameters is the same as on Fig.1, except $R_1=0.4$.}\label{fig2}
\end{figure}

\begin{figure}
  \resizebox{\hsize}{!}{\includegraphics[angle=90]{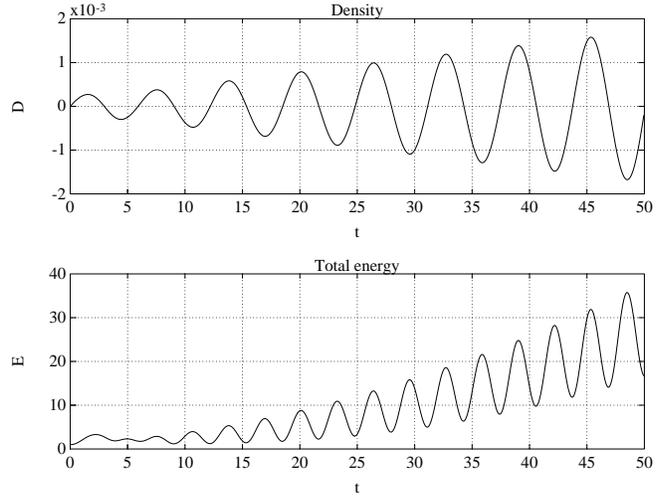}}
  \caption{Shear instability of low-frequency MHD oscillations.
  The plot features temporal
  evolution of the same kind of initial value problem as on
  Fig.1. The only difference is that
  The evolution of parametrically unstable blend of AW and SMW.
  The set of parameters is the same as on Fig.1, except that $a_1=0.1$
  now, making the sign of ${\gamma}^2$ positive.}\label{fig3}
\end{figure}

\begin{enumerate}

\item

The appearance of ``echoing" waves, consisting of repetitive, modulated
bumps of AW and SMW, exchanging energy with the mean flow.
Fig.1 displays an example of such process. From the figure we can
surmise that mutual AW$\rightleftharpoons$SMW transformations still
happen and the resulting mixture of waves ``pulsates" taking
and giving energy from/to the background helical flow! This
process has well-pronounced quasiperiodic nature.

\item

Since coefficients in (22-25) vary periodically it is plausible
to expect that for certain cases the system must also
exhibit some kind of {\it self-parametric} instability. The
term ``self-parametric"
seems appropriate, because it is the consequence of the velocity shear
inherent to the system and forcing on itself (\cite{acr99}). This
kind of instability was first discovered for plain acoustic waves
in 2-D flow patterns of neutral fluids
(\cite{mr99}). Similar sort of instability was found for
Alfv\'en waves in the Paper I. Numerical examination of the (22-25)
allowed us to find similar instabilities for the compressible case.
One example is given on the Fig.2. The values of all parameters (except
$R_1$) here are the same as on the Fig.1, but $R_1=0.4$. The figure shows
that in this case hydromagnetic oscillations amplify exponentially,
extracting energy from the flow. Note that for
the existence of this kind of parametric instability it is {\it
not} necessary to have any periodicity in the background flow,
but it is essential to have periodic time variation of the
wave number vector.

\end{enumerate}

Let us turn our attention, now, to flows with ``sharper" rate of
differential rotation, with $n>1$. In this case ${\gamma}^2>0$
and the temporal evolution of ${\bf k}(t)$'s becomes exponential.
In the absence of dissipation these
flows host rather robust {\it shear instabilities}. The similar
kind of instability was found for acoustic waves
in a 2-D flow (\cite{mr99}) and in the Paper I it was detected
for Alfv\'en waves as well. The example of such instability for the
mixture of AW and SMW is shown on the Fig.3. The parameters are the same as
for the Fig.1, only $a_1=0.1$ (it reverses the sign of the ${\Lambda}^2$).
We see that in this case
the mixture of AW and SMW undergoes rather strong, exponential
enhancement of its amplitude and energy.


\subsection{Low-$\beta$ plasmas}

\begin{figure}
  \resizebox{\hsize}{!}{\includegraphics[angle=90]{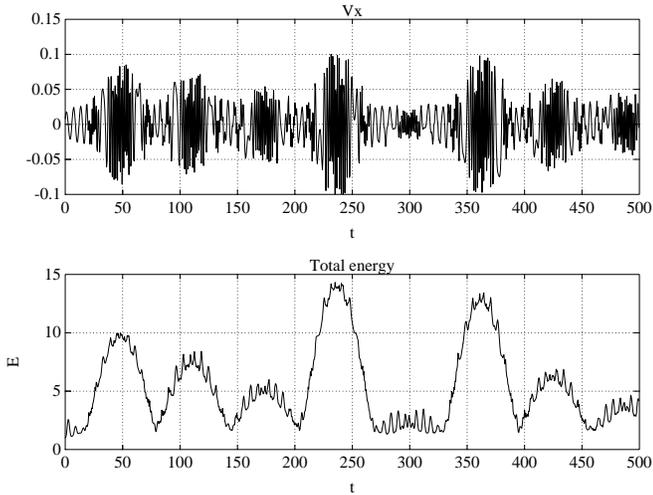}}
  \caption{The temporal evolution of initially excited AW, which became
  partially transformed into FMW and  exhibits quasiperiodic,
  pulsational behavior in low-$\beta$ plasma flow. The set of parameters is:
 $\epsilon=0.1$, $a_1=-0.1$, $a_2=0.1$, $\Sigma=0$, $R_1=0.01$ and $R_2=0.9$.
  Fig.~4 shows time evolution of the velocity perturbation
  $v_x$ and the total energy normalized on its initial value
  $E_{\rm tot}(\tau)/E_{\rm tot}(0)$.}\label{fig4}
\end{figure}

\begin{figure}
  \resizebox{\hsize}{!}{\includegraphics[angle=90]{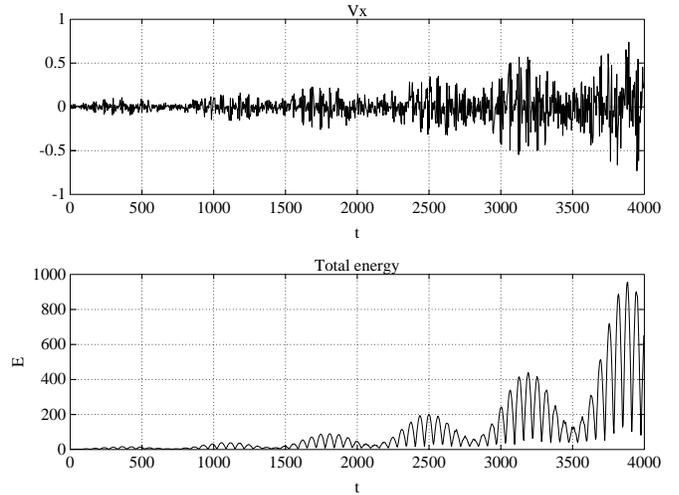}}
  \caption{The parametrically unstable solution. The values of the
  parameters are:  $\epsilon=0.1$, $a_1=-0.1$ $a_2=0.1$, $\Sigma=0$,
  $R_1=0.01$ and $R_2=0.83$.}
\end{figure}

\begin{figure}
  \resizebox{\hsize}{!}{\includegraphics[angle=90]{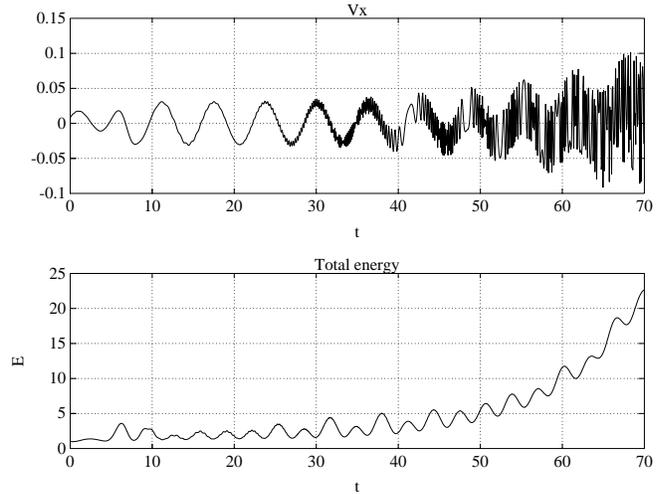}}
  \caption{The shear instability in the $\gamma^2>0$ case. The values of the
  parameters are:  $\epsilon=0.1$,
  $a_1=0.1$ $a_2=0.1$, $\Sigma=0$, $R_1=1$ and $R_2=1$.}
\end{figure}

In this case hydromagnetic oscillations offer quite a different
picture. Namely, the SMW mode is decoupled from the other two MHD
modes---its dispersion curve runs well below the Alfv\'en
dispersion ``horizontal" and it is not coupled with other two
(AW and FMW) modes. The FMW, on the
other hand is coupled with the AW. When rotation is
absent and there is only parallel outflow the coupling ensures
linear transformation of the AW into the FMW (see as an example
Fig.6a-d in \cite{rpm00}).

In the helical flow, getting certain ``input" of initial AW and/or
FMW oscillations, these waves will keep transforming into each
other. Besides the mixture of waves might exhibit the same kind
of ``echoing" and unstable behavior as it was seen in high-$\beta$
plasmas. Numerical simulations support this expectation. Making values of
$a_1$ and $a_2$ nonzero ($a_1=-0.1$ and
$a_2=0.1$, so that $\gamma^2<0$) and using following parameters:
$R_1=0.01$ and $R_2=0.9$  we see (Fig.4) the appearance of interesting patterns
of ``echoing" solutions with quasiperiodic variability of the
perturbation total energy\footnote{Note that here, as elsewhere on the plots of this
paper, the total energy is normalized on its initial value.}.

We find that parametric instabilities are also characteristic to
low-$\beta$ plasmas. The example is given on Fig.5.
The parametric nature of the instability is apparent from the
remarkable narrowness of the range of parameter values, where the
instability is present. Note that the set of parameters used for
Fig.5 is the same as for Fig.4 except $R_2=0.83$. If one takes the
value of $R_2$ less by $0.01$, then the instability disappears and
the system again displays the ``echoing behaviour.

When the
differential rotation is characterized by $n>1$ profile
(${\gamma}^2>0$) the system shows strong exponential
shear instability. The corresponding plots are given on Fig.6.



\subsection{The case of $\beta \simeq 1$}

\begin{figure}
  \resizebox{\hsize}{!}{\includegraphics[angle=90]{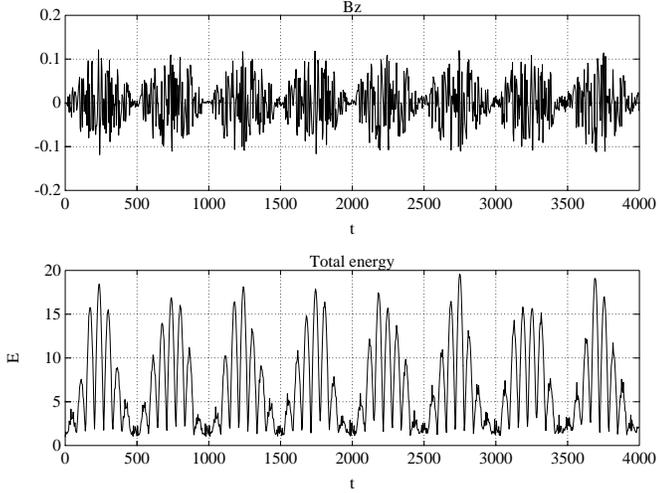}}
  \caption{The temporal evolution of the wave blend that exhibits
quasiperiodic, pulsating behavior in $\epsilon=1$ plasma flows.
The set of parameters is: $a_1=-0.1$ $a_2=0.1$, $\Sigma=0$,
$R_1=0.1$ and $R_2=0.85$. The plots show time evolution of the
magnetic field $B_x$ perturbations and the normalized energy
$E_{\rm tot}(\tau)/E_{\rm tot}(0)$.}\label{fig7}
\end{figure}

\begin{figure}
  \resizebox{\hsize}{!}{\includegraphics[angle=90]{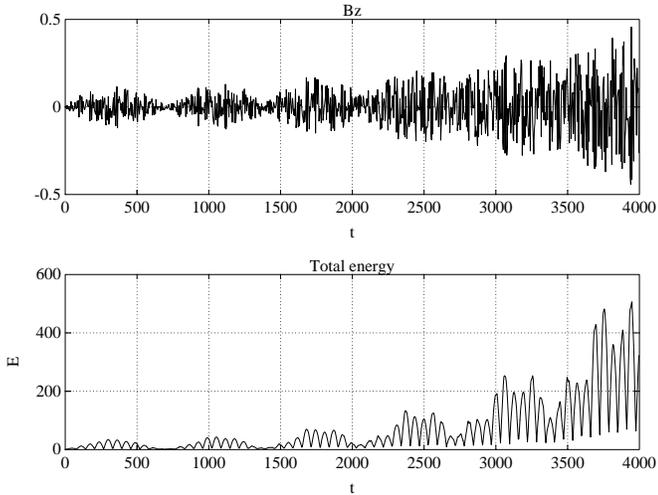}}
  \caption{The evolution of parametrically unstable blend of waves. The set of
parameters is: $a_1=-0.1$ $a_2=0.1$, $\Sigma=0$, $R_1=0.1$ and
$R_2=0.8005$. Note that the range of $R_2$ when the evolution of
the wave mixture is parametrically unstable is very narrow:
[0.8001; 0.8011].}\label{fig8}
\end{figure}

\begin{figure}
  \resizebox{\hsize}{!}{\includegraphics[angle=90]{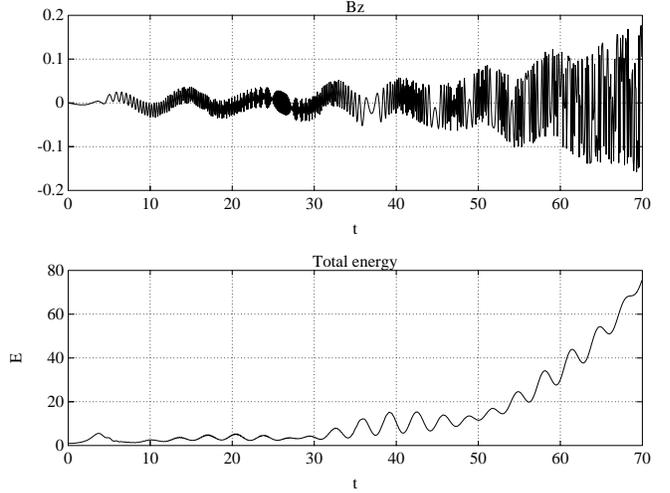}}
  \caption{The shear instability in the $\lambda^2>0$  case. The set of
parameters is: $a_1=0.1$ $a_2=0.1$, $\Sigma=0$, $R_1=1$ and
$R_2=1$. }\label{fig9}
\end{figure}

From the studies of parallel flows we know that in terms of wave
couplings and mutual transformations this is the most complex case:
all MHD wave modes are coupled and may transform into
each other. In helical flows the presence of wave
transformations may be less visible, overshadowed (when
$\gamma^2<0$) by quasiperiodic modulation of waves appearing as
repetitive, (``echoing")  bundles of mixed AW, SMW and FMW modes
(see Fig.7); or by self-parametric instability (see
Fig.8). While when $\gamma^2>0$ the waves are unstable in the
similar way (see Fig.9) as in previously considered high-$\beta$
and low-$\beta$ cases.

\section{Conclusion}

The goal of this work was to find out whether the exotic SINP
originally found for 2-D kinematically complex velocity patterns
of neutral fluids (\cite{mr99})
and for 3-D helical flows of magnetized conducting
fluids in the incompressible limit, appear also for
the full spectra of MHD waves sustained by compressible MHD
medium. We found that this is indeed the case! Therefore we can now firmly
claim that the range of the SINP, typical for the flows
of complicated, helical nature, is broad. These SINP persist to show up
both in the incompressible and compressible cases. They are
present in flows with arbitrary values of plasma-$\beta$.

In relatively mildly sheared flows (with $n<1$, i.e., including
rigidly rotating systems) MHD modes appear to be rather stable
exhibiting either ``echoing" pulsational behaviour or relatively
long-time-scaled parametric instabilities. One can expect that in
helical flows of this nature, especially in
well-beamed, or well-collimated flows such as jets, shear flow
effects are not likely to lead to disruptive instabilities.
Instead, through quasiperiodic interchange of energy with the mean flow,
they might tend to exhibit certain modes of quasi stable and quasiperiodic
structuring both in the space and in time.

In more strongly sheared systems ($n>1$ including Keplerian
rotation) waves within swirling flows become subject to
potentially very fast-growing shear instabilities, which would
either lead to the disruption of ``parent" flow patterns or to the
development of the MHD turbulence with subsequent
phase transition to turbulent rotational flow systems. We have to
bear in mind that in this case the exponential growth of the $|\bf
k(t)|$ inevitably makes the spatial length-scales of perturbations
smaller and smaller. It implies that the effects of the viscous
decay and/or magnetic diffusion, neglected while we consider the
MHD flow as an  ``ideal" one, must sooner or later become
important and lead to the dissipation of the energy gained by the
exponentially increasing waves into the heat. It can be argued
that in accretion-ejection systems, where the rotational law seems
to be quasikeplerian, these instabilities may account for the
transition to turbulence in accreted plasma flows. Alternatively,
this process might lead to effective ``self-heating" of
these flows, when the energy acquired by waves from the flows
through the agency of the shear instability would eventually
transform into heat via diffusion.

Speaking about swirling astrophysical flows we are keen to use the
term {\it ``cosmic tornado"} for this class of flows, because they are
reminiscent of powerful and dangerous tornados in the Earth's
atmosphere. Recently such structures, called {\it solar tornados},
were identified in the polar regions of the solar atmosphere (both
on the limb and the disk) by SOHO-CDS observations
(\cite{pm98}). Another class of cosmic tornados are, probably, stellar jets,
because recent observational results (\cite{b02, d00}) seem to confirm the
predictions of various magnetocentrifugal jet acceleration models
(\cite{snos95, fp95, c97, lhf99, kp00}) about the presence of a swirling
motion within the jets. The third class of
astrophysical flows with possible presence of tornado-like motion are
{\it accretion columns} -- magnetically channeled
shear flows of plasma to a neutron star's (or a white dwarf's) magnetic
pole. The
infalling matter is decelerated approaching the star surface, but
it can also form a shock high above the star's surface
(\cite{hp98}). Originally the formation of rotational accretion
columns was considered in the astrophysical fluid dynamics context
(\cite{c78}), while now accretion columns associated with X-ray pulsars and
cataclysmic binaries are modeled either as thin slabs or tall
columns of infalling matter. It seems reasonable to surmise that accretion
columns comprise 3-D swirling plasma flows.

Bearing in mind these perspectives we should stress that the
results of this paper are quite general and (deliberately) not
adjusted to either of three kinds of swirling astrophysical
flows. For all these classes of `cosmic tornados' the level of our
factual knowledge about their basic kinematic features is still inadequate
for building of any credible concrete models. However we hope that these
future models, based on the gained data and implying real-space sophisticated
simulations, will show how the generic processes disclosed in
this paper and the Paper I might influence the overall dynamics of
real `cosmic tornados'.

\section{Acknowledgements}

The authors are grateful to Swadesh Mahajan and Stefaan Poedts for
valuable discussions. Andria Rogava and Zaza Osmanov are grateful
to the International Centre for Theoretical Physics for supporting
them, in part, through the Regular Associate Membership Award and
the Young Collaborator grant, respectively. Andria Rogava is
grateful to the Universit\'a degli Studi di Torino for supporting
him, in part, through the {\it Assegno di Ricerca e
Collaborazione}.

\end{document}